\documentclass[twocolumn,preprintnumbers,aps]{revtex4-1}
\usepackage{amsmath}
\usepackage{amssymb} 
\usepackage{color}
\usepackage{float}
\usepackage{sistyle}
\usepackage{subfigure}  
\usepackage{verbatim}   
\usepackage{graphicx}
\usepackage{dcolumn}
\usepackage{bm}
\usepackage[citecolor=blue,filecolor=blue,colorlinks=true,urlcolor=blue]{hyperref}
\usepackage{ulem}
\usepackage{lineno}
\usepackage{silence}
\newcommand{\BAN}{\ensuremath{B_{1g}\,}}
\newcommand{\BN}{\ensuremath{B_{2g}\,}}

\newcommand{\Ts}{\ensuremath{T^{\ast}\,}}
\newcommand{\Tc}{\ensuremath{T_{\rm c}\,}}
\newcommand{\cm}{\ensuremath{{\rm cm}^{-1}}}

\graphicspath{{images/}{../images/}}

\begin{document}

\title{Universal relationship between the energy scales of the pseudogap phase, the superconducting state and the charge density wave order in copper oxide superconductors}

\author{B. Loret$^1$, N. Auvray$^1$, G. D. Gu$^2$, A. Forget$^3$, D. Colson$^3$, M. Cazayous$^1$, Y. Gallais$^1$, I. Paul$^1$, M. Civelli $^4$ and A. Sacuto$^1$${^\ast}$}
\affiliation{$^1$ Universit\'e de Paris, Laboratoire Mat\'eriaux et Ph\'enom$\grave{e}$nes Quantiques (UMR 7162 CNRS),Bat.Condorcet, 75205 Paris Cedex 13, France\\
$^2$ Matter Physics and Materials Science, Brookhaven National Laboratory (BNL), Upton, NY 11973, USA,\\
$^3$ Universit\'e  Paris-Saclay, CEA, CNRS, SPEC, 91191, Gif-sur-Yvette, France\\
$^4$ Universit\'e Paris-Saclay, Laboratoire de Physique des Solides, CNRS, 91405 Orsay Cedex, France\\}

\date{\today}

\begin{abstract}
We report the hole doping dependencies of the pseudogap phase energy scale, $2\Delta_{\rm PG}$, the anti-nodal (nodal) superconducting energy scales $2\Delta^{AN}_{\rm SC}$ ($2\Delta^{N}_{\rm SC}$) and the charge density wave energy scale, $2\Delta_{\rm CDW}$ extracted from the electronic Raman responses of several copper oxide families. We show 
for all the cuprates studied, that the three energy scales $ 2\Delta_{\rm PG}$, $2\Delta^{AN}_{\rm SC}$ and $2\Delta_{\rm CDW}$ display the same decreasing monotonic behavior with doping. In particular, $2\Delta^{AN}_{\rm SC}$ and $2\Delta_{\rm CDW}$ have nearly equal values. This suggests an universal scenario in which  $2\Delta_{\rm PG}$, $2\Delta^{AN}_{\rm SC}$ and $2\Delta_{\rm CDW}$ are governed by common microscopic interactions that become relevant well above the superconducting transition at \Tc. This is to be contrasted with the behavior of the nodal superconducting energy scale $2\Delta^N_{\rm SC}$, which tracks the doping dependence of \Tc and, hence, seems to be controlled by different interactions.

\end{abstract}

\maketitle

\section{Introduction}
The copper oxide (cuprate) superconductors are materials with an extremely rich Temperature-doping ($T-p$) phase diagram. By decreasing $T$, a mysterious phase is established, the pseudogap (PG) which manifests itself by the suppression of low energy electronic states. By further reducing temperature, a charge density wave (CDW) order settles down. Finally, at lower $T$ superconductivity (SC) arises\cite{keimer2015}. The overriding question that remains unanswered since the discovery of superconductivity in cuprates by Bednorz and Muller in 1986 \cite{Bednorz86} is what are the underlying quantum electronic orders that control the ($T-p$) cuprate phase diagram? 
This question leads to other recurring questions such as the following. Why is the doping dependencies of the superconducting transition temperature $\Tc(p)$ and the charge density wave transition temperature $T_{\rm CDW}(p)$ dome-like? Why instead the pseudogap temperature $\Ts(p)$ decreases linearly as $p$ increases? In order to address these key questions it is essential to identify the relation between the transition temperatures $\Tc(p)$, $T_{\rm CDW}(p)$, $\Ts(p)$ and the corresponding energy scales, $2\Delta^{AN}_{\rm SC}$ ($2\Delta^{N}_{\rm SC}$), $2\Delta_{\rm CDW}$, $2\Delta_{\rm PG}$. Here AN and N refer to anti-nodal and nodal regions corresponding to the principal axes and the diagonal of the first Brillouin zone (BZ) respectively. For mean field type second order phase transitions, such as superconducting transitions in conventional systems \cite{Bardeen57}, the transition temperature is proportional to the associated energy scale of the order parameter, such as the gap value. As we discuss in detail below, this is not the case either for the superconducting or the charge density wave transitions of the cuprates, which emphasizes their unconventional nature. 

The scope of this article is to determine the above mentioned energy scales of various cuprate families, extending our previous work on tri-layered HgBa$_2$Ca$_2$Cu$_3$O$_{8+\delta}$\cite{Loret2019}. The goal here is to show that their relation with the transition temperatures and their behaviour as a function of doping are universal features of the cuprate phase diagram. Our findings also provide important clues on the relationship between the different quantum electronic orders. 

Our study is based on the electronic Raman spectroscopy (ERS), which is a very effective probe to track the energy scales  of the superconducting gap, the pseudogap \cite{Blumberg1997,Opel2000,LeTacon2006,Devereaux2007,Blanc2010,benhabib15,Loret2017a,Loret2018} or more recently the charge density wave gap \cite{Loret2019}. Since ERS is a two photon scattering process, by controlling the incoming and outgoing photon polarizations, one can selectively probe different regions of the BZ. Thus, in the \BAN geometry the Raman form factor is $( \cos k_x - \cos k_y)^2$ and it predominantly probes the anti-nodal region. Here ${\bf k}$ is the wave vector of the excited electron. Likewise, in the \BN geometry the Raman form factor is $\sin^2 k_x \sin^2 k_y$ and it probes mostly the nodal region. This form factor induced momentum space selectivity is particularly useful for studying the cuprates, since it is well known that the electronic properties in the anti-nodal and nodal regions are quite different\cite{Norman03}. Thus, both the superconducting gap and the pseudogap energy scale are maximal around the anti-node, and minimal around the nodal region. Furthermore, since the energy gaps from pseudogap is minimal in the nodal region, we find that any additional loss of low-energy spectral weight due to the formation of the charge density wave is readily detected in the \BN geometry, and not in the \BAN geometry where the signal of the charge density wave is masked by the pseudogap. Yet another useful aspect of ERS is that it is a frequency-resolved, but, (form factor weighted) momentum averaged probe. Therefore it is quite sensitive to gap opening due to a short range order, as is the case of the charge density wave in the cuprates in the absence of magnetic field. This is because any spatial variation of the ordering wave vector in a short range order does not blur the energy gap feature in frequency space if the probe is frequency resolved \footnote{Provided the energy scale variations are not too large over different patches in which an electronic order sets in}. 
The ERS measurements were performed on four distinct cuprates: HgBa$_2$CuO$_{4+\delta}$ (Hg-1201), YBa$_2$Cu$_{3}$O$_{6+\delta}$ (Y-123), Bi$_2$Sr$_2$CaCu$_{2}$O$_{8+\delta}$ (Bi-2212) and HgBa$_2$Ca$_2$Cu$_3$O$_{8+\delta}$ (Hg-1223). In our previous study\cite{Loret2019}, we determined the doping dependence of the energy scales of the Hg-1223 compound and the doping trend of the CDW energy scale of Y-123 compounds. Here, among the new results, we were able to identify the CDW energy scale and the doping dependence of the PG energy scale in Bi-2212 compound. We also succeeded to follow the doping dependence of the CDW energy scale in Hg-1201 compound by reinterpreting the Raman data obtained from other group \cite{Li2013}. This allows us to obtain an universal picture of the doping evolution of the energy scales over several cuprates. 
We find that $2\Delta_{\rm PG}$, $2\Delta^{AN}_{\rm SC}$ and $2\Delta_{\rm CDW}$ have the same doping dependency as $\Ts (p)$, they decrease linearly with doping. This suggests that they are all driven by the same microscopic mechanism. We also find that the $2\Delta_{\rm PG}$ is approximately twice larger than $2\Delta^{AN}_{\rm SC}$ and $2\Delta_{\rm CDW}$ which are instead very close to each other. This suggests first, that $2\Delta_{\rm PG}$ scale ( detected here) cannot be ascribed to SC fluctuations \cite{Emery1995} (since it is too far from the SC gap scale $2\Delta^{AN}_{\rm SC}$). Secondly and most importantly, this indicates that the SC and CDW orders are intimately connected (since $2\Delta^{AN}_{\rm SC} \approx 2\Delta_{\rm CDW}$) and deserve to be investigated in the light of recent theoretical models such as composite or intertwined orders \cite{Efetov2013,Sachdev13,Fradkin2014,Wang2015,Caprara2017,Chakraborty2019}.

\section {Anti-nodal Superconducting and Pseudogap energy scales} 

Our first goal is to show how we can detect and define the energy scales of the anti-nodal part of the SC gap and the PG phase in cuprates. As an illustration, we present the \BAN Raman responses of the Hg-1223 (UD 117) and Bi-2212 (UD 75) compounds. The number in brackets corresponds to the \Tc value. As mentioned above, the \BAN  geometry in Raman spectroscopy gives us access to the anti-nodal part of the BZ. The details of the ERS experimental procedure is given in Appendix A. Details on the crystals growth, the \Tc values and doping are given in Appendix B. 

\begin{figure}[ht!]
\begin{center}
\includegraphics[scale=0.47]{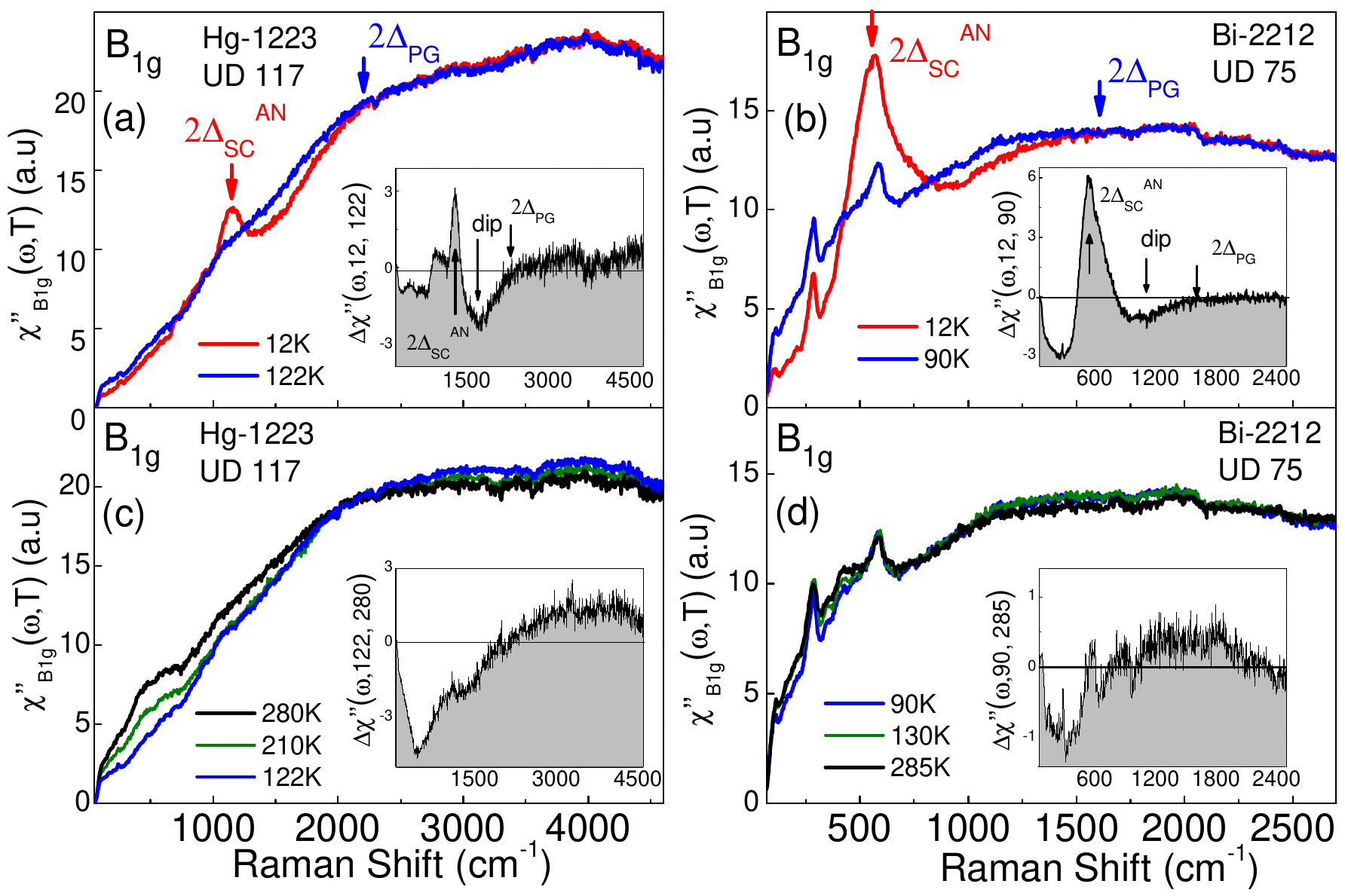}
\caption{(Color online). For $T\leq\Tc$, \BAN (anti-nodal) Raman response functions of (a) an under-doped Hg-1223 (UD 117, $p\approx 0.12$) and (b) Bi-2212 (UD 75, $p\approx 0.11$) single crystal; (c) and (d) For $T\geq\Tc$. The SC energy scale $2\Delta_{\rm SC}$ and the PG energy scale $2\Delta_{\rm PG}$ are indicated by red and blue arrow respectively. In the insets of panels (a) and (b), the subtracted Raman response (defined in the text) underlines the peak-dip structure.  In the insets of panels (c) and (d), the subtracted Raman response (defined in the text), points out the spectral weight transfer induced by the pseudogap phase in the normal state. The shaded area allows by contrast, a better visualization of the peak-dip structure in the SC state and the spectral weight transfer in the normal state.} 
\label{fig:1}
\end{center}\vspace{-7mm}
\end{figure}
In Figure 1 (a) and (b), the SC Raman responses of Hg-1223 (UD 117) and Bi-2212 (UD 75) at 12 K, exhibit a well defined pair breaking peak, $2\Delta{\rm ^{AN}}_{\rm SC}$, at approximately two times the energy of the SC gap measured by tunneling and angular resolved photo-emission spectroscopy (ARPES). It corresponds to the maximum amplitude of the $d$-wave SC gap probed at the anti-nodes.  It is located around 1500 \cm and 570 \cm for Hg-1223 (UD 117) and Bi-2212 (UD 75) respectively and marked by a red arrow.  The $2\Delta{\rm ^{AN}}_{\rm SC}$ peak is associated on its high energy side with a dip in the electronic continuum. The dip is revealed by comparing the SC (at 12 K) and the normal Raman responses just above \Tc (122 K for Hg-1223 (UD 117) and 90 K for Bi-2212 (UD 75)). In previous works, we showed that this peak-dip structure detected in the superconducting \BAN Raman response results from the interplay between the PG and the SC gap \cite{Loret2016,Loret2017a,Loret2018}.
The peak-dip structure is emphasized by the subtracted Raman responses $\Delta\chi^{\prime\prime}_{\BAN} (\omega,12K, 122K)=\chi^{\prime \prime}_{\BAN}(\omega, 12K)-\chi^{\prime \prime}_{\BAN}(\omega, 122K)$ and $\Delta\chi^{\prime \prime}_{\BAN} (\omega,12K, 90K)$ for Hg-1223 (UD 117) and Bi-2212 (UD 75) respectively (see insets of panels (a) and (b)). We established (i) the peak-dip structure is only detected when the pseudogap exists \cite{Loret2016,Loret2017a,Loret2018} and (ii) it can be smoothly connected to the loss of spectral weight related to the PG above \Tc (see Appendix C). Above \Tc, as the temperature is lowered, see Fig.1 (c) and (d), we observe simultaneously a loss and a slightly increase of spectral weight of the electronic background below and above 2000 \cm  and 1000 \cm for respectively Hg-1223 (UD 117) and Bi-2212 (UD 75). This is due to a quasi-particles spectral weight transfer from low to high frequency which characterizes the pseudogap phase. This is underlined by the subtracted Raman responses $\Delta\chi^{\prime \prime}_{\BAN} (\omega,122K, 280K)$ (for Hg-1223 (UD 117)) and $\Delta\chi^{\prime \prime}_{\BAN} (\omega,90K, 285K$) (for Bi-2212 (UD 75)) which signal the loss and the increase of the spectral weight in the negative and positive part of the spectra respectively, as shown in the insets of the (c) and (d) panels. In the positive part of the spectra, it is hard to accurately define an energy scale for the PG in the normal state, because the hump is almost flat on a large frequency range (2500-4500 \cm) and (800-2000 \cm) for Hg-1223 (UD 117) and Bi-2212 (UD 75) respectively. On the other hand, the energy of the dip end of the SC \BAN response is more easily detectable and since it corresponds to the Raman signature of the PG in the SC state \cite{Loret2016,Loret2017a,Loret2018,Loret2019}, we defined it as the PG energy scale, $2\Delta_{\rm PG}$, marked by a blue arrow in Fig.~\ref{fig:1}. 
$2\Delta_{\rm PG}$ corresponds to the high energy pseudogap detected in the tunneling and ARPES measurements \cite{Dipasupil2002,McElroy05,Campuzano99} and discussed in Ref.\cite{Lee06}. The $2\Delta{\rm ^{AN}}_{\rm SC}$ and $2\Delta_{\rm PG}$ scales for the Y-123 and Hg-1201 compounds were obtained using the same method (analyzing the peak-dip structure in the SC Raman response \cite{Loret2017a}). All the energy scales that we have measured by ERS for the various cuprates are shown in Fig.~\ref{fig:3}. We have also defined the pseudogap temperature \Ts by the temperature at which the transfer of spectral weight from low to high energies detected in the \BAN Raman response ceases. The \Ts values that we obtained, are in agreement with those obtained from other techniques, and are reported in Fig.~\ref{fig:3}. 

\section {Nodal Superconducting and Charge Density Wave Energy Scales} 

We investigate now, the energy scales of the nodal SC gap and the CDW order. The \BN geometry in Raman spectroscopy, allows us to capture the electronic states located in the nodal region of the first BZ, it is therefore well adapted to study the nodal component of the superconducting gap. On the other hand, the parts of the Fermi surface which are expected to be the most affected by the CDW order are located in between the nodal and the anti-nodal regions \cite{Comin2016}, meaning that they are potentially accessible by both the \BAN or \BN geometry.
However, since $T_{\rm CDW}$  is below \Ts, the CDW signal close to the anti-nodes may be affected or masked by the pseudogap spectral weight loss. It is therefore more adapted to look for the CDW signal close to the nodes, where the pseudogap effect is known to be minimal. In our previous investigations, we have mostly investigated the CDW signal in Hg-based compounds\cite{Loret2019}. Here, we show that we are also able to detect the CDW signal also in Y-123 and Bi-2212 compounds, despite the presence of few phonon modes which are not detected in Hg-based compounds. So, we have studied the \BN Raman response of the four following cuprates: Hg-1223 (UD 117) , Hg-1201 (UD 72) and Y-123 (UD 54) and Bi-2212 (UD 75). The details of the crystal growths and characterizations of Hg-1201 and Y-123 can be found in Refs. \cite{Legros2019,Alloul2010}.

\begin{figure}[ht!]
\begin{center}
\includegraphics[scale=0.39]{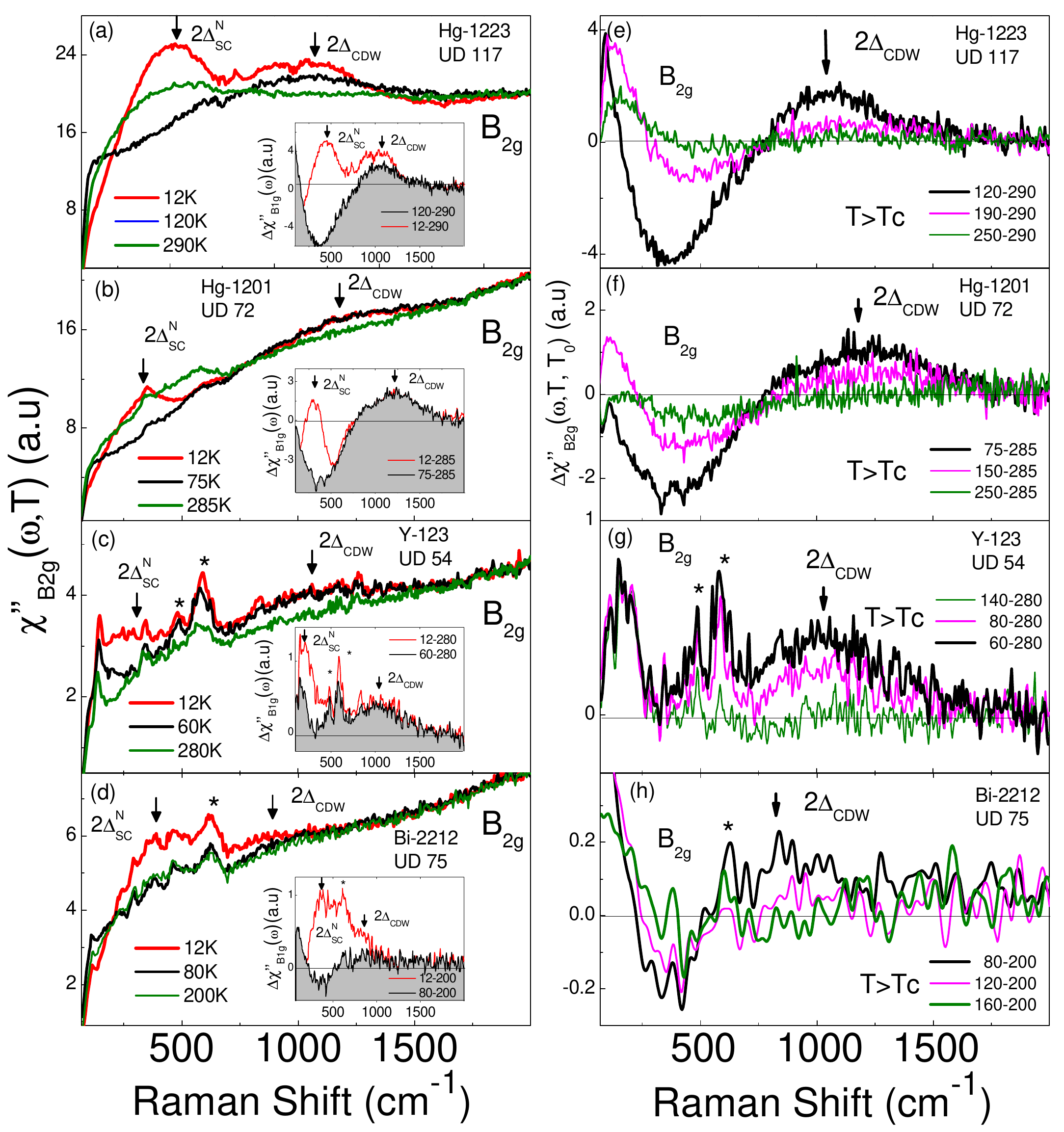}
\caption{(Color online). \BN (Nodal) Raman response functions of Hg-1223 (UD 117, $p\approx 0.12$) , Hg-1201 (UD 72, $p\approx 0.11$) and Y-123 (UD 54, $p\approx 0.08$) and Bi-2212 (UD 75, $p\approx 0.11$) compounds. (a)-(d) for selected temperatures above and below \Tc. In the insets, we plot the subtraction between the Raman responses measured below (red curve) and  just above (black curve) \Tc and the ones measured at $T_0$. $T_{0}$ = 290 K, 285 K, 280 K, 200 K for Hg-1223, Hg-1201, Y-123 and Bi-2212 respectively. The shaded area allows a better visualization of the CDW signal. (e)-(h) Subtracted Raman responses between selected temperature above \Tc  and $T_{0}$. The peaks labeled by a star in the Y-123 and Bi-2212 are phonon modes induced by oxygen disorder \cite {Bakr13,Benhabib2015}. Note that since the CDW signal is intrinsically weak in Bi-2212 compound, we applied a slight filtering by using Fourier transform to improve the signal noise ratio of the Raman response in panel (h).}
\label{fig:2}
\end{center}\vspace{-7mm}
\end{figure}
\begin{figure*}
\begin{center}
\includegraphics[scale=0.9]{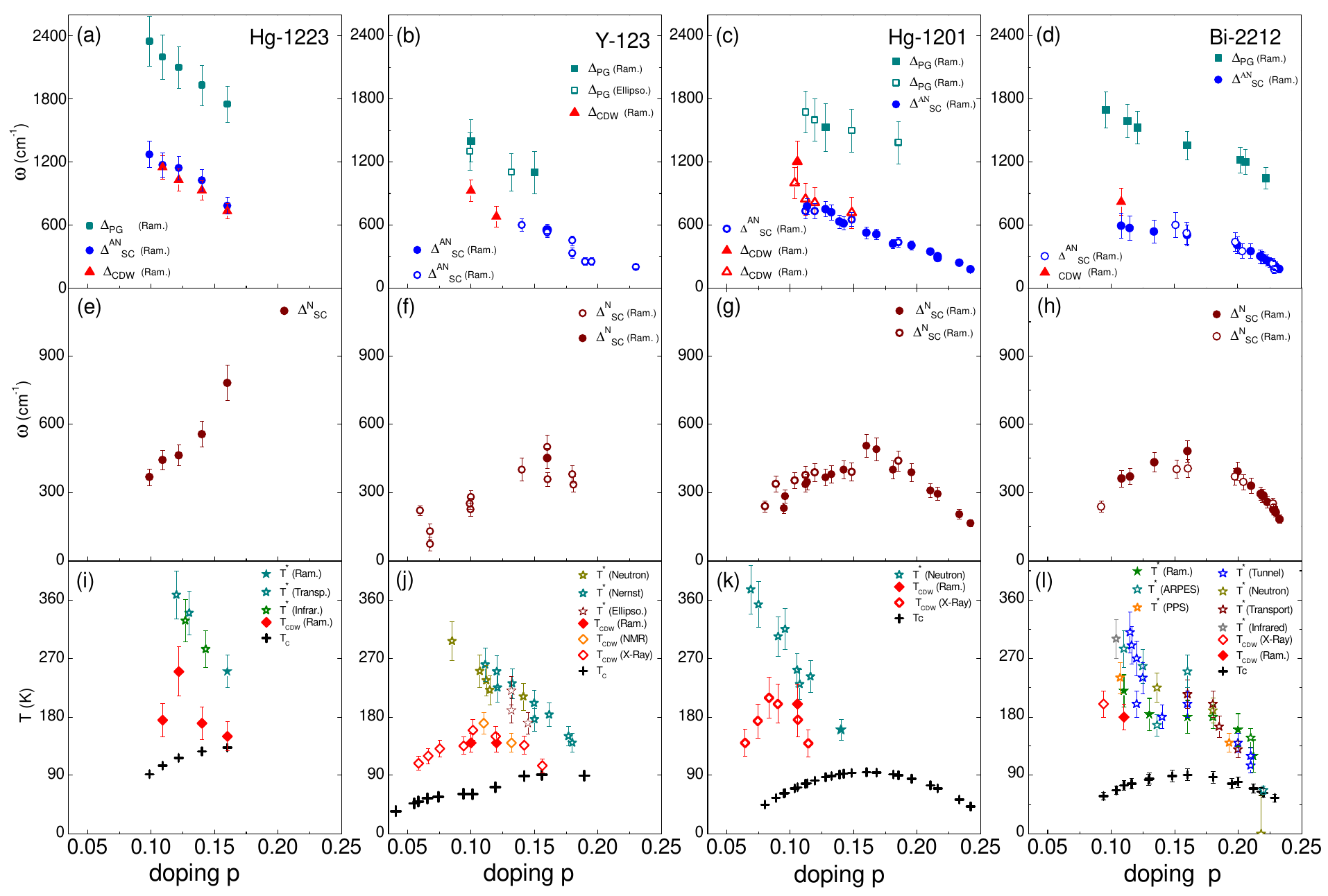}
\caption{(Color online). Universal doping dependencies of the pseudogap, the anti-nodal superconducting and the charge density wave energy scales, (respectively $2 \Delta_{\rm PG} (p)$, $2 \Delta^{AN}_{\rm SC} (p)$ and $2\Delta_{\rm CDW}(p)$) over four cuprates systems: (a) Hg-1223, (b) Y-123, (c) Hg-1201 and (d) Bi-2212 cuprates. Panels (e), (f), (g), (h) show the doping dependence of the nodal superconducting energy scale, $2 \Delta^{N}_{\rm SC}(p)$, for Hg-1223, Y-123, Hg-1201 and Bi-2212 respectively. Panels (i), (j), (k), (l) display the doping dependence of the relevant transition temperatures: the pseudogap \Ts, the superconducting \Tc and the charge density wave $T_{\rm CDW}$ for Hg-1223, Y-123, Hg-1201 and Bi-2212 respectively. The symbols filled in, correspond to our Raman data. Our data on Y-123, Hg-1201 and Bi-2212 are supplemented by Raman measurements from other groups (designated by empty symbols), see ref.[37]. } 
\label{fig:3}
\end{center}\vspace{-7mm}
\end{figure*}

Below \Tc (red curves in the (a) to (d) panels of Fig.2), we detect two distinct features arrowed $2\Delta{\rm ^N}_{\rm SC}$ and $2 \Delta_{\rm CDW}$ in the Raman spectra of Hg-1223, Hg-1201, Y-123 and Bi-2212. These features are highlighted by the subtracted Raman responses $\Delta\chi^{\prime \prime}_{\BN} (\omega,T=12K,T_0)$ obtained from the difference between the Raman responses taken at $T$=12 K and the ones measured at $T_0>$\Tc (see red curves in the insets of panels (a) to (d)). $T_0$ is defined in the caption of Fig.2. The extra narrow features labeled by a star in Fig.2 correspond to phonon lines (see caption of Fig.2). The $2\Delta{\rm ^N}_{\rm SC}$ peak located at 470 \cm for Hg-1223, 300 \cm for Hg-1201, 250 \cm for Y-123 and 372 \cm for Bi-2212, disappears above \Tc. This feature is assigned to the well known nodal component of the $d$-wave SC gap and already extensively studied in Y-123, Bi-2212 and Hg-1201 \cite{Opel2000,LeTacon2006,Guyard2008a,Munnikes2011}. On the contrary, $2 \Delta_{\rm CDW}$ located at higher frequency than $2\Delta{\rm ^N}_{\rm SC}$, persists above \Tc (see black curves in the insets of panels (a) to (d)) and exhibits a maximum around 1000 \cm for Hg-1223 (UD 117), 1150 \cm for Hg-1201 (UD 72), 1000 \cm for Y-123 (UD 54) and 850 \cm for Bi-2212 (UD 75). We define the position of the maximum of the hump as the CDW energy scale, 2$\Delta_{CDW}$ which has been first identified in ref.\cite{Loret2019} for Hg-1223. All these energy scales are reported in Fig.~\ref{fig:3}. 
By raising the temperature above \Tc, the CDW hump progressively decreases in intensity. This is pointed out by looking at the subtracted Raman responses $\Delta\chi^{\prime \prime}_{\BN} (\omega, T>\Tc,T_0)$  of Hg-1223, Hg-1201, Y-123 and Bi-2212 (panels (e)-(h)). They exhibit a dip-hump structure characteristic of the CDW spectral weight transfer that disappears with increasing $T$. We have defined $T_{\rm CDW}$ as the temperature at  which the integrated Raman intensity of the CDW hump vanishes. The $T_{\rm CDW}$ values of Hg-1223, Hg-1201 and Y-123 and Bi-2212 (UD 75) are reported in Fig.~\ref{fig:3} and are in agreement with those obtained from other techniques, see Fig.3. 
At this stage, it is important to notice that although $T_{\rm CDW}$ has been extensively mapped out in several cuprates, only very few data have been reported on the $2 \Delta_{\rm CDW}$ energy scale. One can then legitimately wonder whether the CDW signal we observed in \BN geometry is also detected by other spectroscopic techniques in the Bi-2212 compound. We can effectively find that polarized pump probe measurements have reported such signals at nearly the same energy \cite{Toda2014}. Additionally, ARPES measurements report a gap opening in the nodal region (\BN geometry) below \Ts which could be related to a CDW order \cite{Kaminski2015}. 

\section {Universal doping dependence of the energy Scales in several cuprates and its relationship on the cuprates phase diagram.} 
Having showed how to extract and identify the energy scales of the SC state, the PG phase and the CDW order from the electronic Raman response, our objective is to track their doping dependencies for various families of cuprates that we studied, and see if there exists some common trends and how they can eventually be connected on the cuprate phase diagram. By way of illustration, we have reported in Appendix D, the Raman responses of the Hg-1223 compound for several doping levels from which we extracted the energy scales: $2\Delta{\rm ^{AN}}_{\rm SC}$, $2\Delta_{\rm PG}$, $2\Delta{\rm ^N}_{\rm SC}$ and $2 \Delta_{\rm CDW}$. The doping dependence of these four energy scales are shown in panels (a) and (e) for Hg-1223, (b) and (f) for Y-123, (c), (g) for Hg-1201 and (d), (h) for Bi-2212 \footnote{In Y-123, $2 \Delta_{\rm PG} (p)$ values were extracted from ellipsometry \cite{Bernhard2008,Dubroka11} and Raman \cite{Loret2017} measurements, $2\Delta^{AN}_{\rm SC} (p)$ from Raman \cite{Opel2000,Masui2003} and $2\Delta_{\rm CDW}(p)$ from Raman \cite{Loret2019}. In Hg-1201, $2 \Delta_{\rm PG}(p)$ and $2 \Delta^{AN}_{\rm SC} (p)$ values were extracted from the \BAN SC Raman spectra \cite{LeTacon2006,Guyard2008a,Guyard2008,Li2013,Loret2017,Loret2019} as detailed in first section. We extracted the $2\Delta_{\rm CDW}(p)$ from the \BN Raman spectra \cite{Li2013,Loret2019}. Note that, we have re-interpreted the data of Li et al.\cite{Li2013} in the light of our recent works\cite{Loret2019}. In Bi-2212, the $2\Delta_{\rm PG}(p)$ values come from Raman \cite{Loret2017}. The $2 \Delta^{AN}_{\rm SC}(p)$ values were extracted from Raman \cite{Venturini2002,Munnikes2011,benhabib15,Loret2017,Loret2018}. In Y-123, Hg-1201 and Bi-2212 the $2 \Delta^{N}_{\rm SC} (p)$ values were extracted from Raman data stemming from refs. \cite{Opel2000,Sugai03,Munnikes2011}, refs. \cite{Gallais2006,LeTacon2006,Guyard2008a,Guyard2008,Li2013,Loret2019} and refs. \cite{Venturini02,Munnikes2011,benhabib15} respectively. In Hg-1223, \Ts values were extracted from transport \cite{Carrington1994,Julien1996}, infrared \cite{McGuire2000} and Raman \cite{Loret2019},  $T_{\rm CDW}$ and \Tc from Raman and magnetic susceptibility \cite{Loret2019}. In Y-123, \Ts values were extracted from ellipsometry \cite{Bernhard2008}, transport \cite{Daou2010} and neutron \cite{Sidis2013} data, $T_{\rm CDW}$ and \Tc from X-ray and transport \cite{Blanco-Canosa14,Huecker2014,Comin2016,Arpaia2019} and nuclear magnetic resonance (NMR)\cite{Wu2015}. In Hg-1201, \Ts values were extracted  from Raman \cite{Guyard2008}, Neutron \cite{Li2008,Baledent11,Li2011} and transport \cite{Barisic2013}, $T_{\rm CDW}$ and \Tc from X-ray and transport \cite{Tabis2017}.  In Bi-2212, \Ts and \Tc values were extracted from transport \cite{Watanabe97,Usui2014}, infrared \cite{Hwang2004}, tunneling \cite{Dipasupil2002,Ozyuzer2002} ARPES \cite{Vishik12,Kaminski2015}, neutron\cite{Mangin-Thro2014}, pump probre spectroscopy (PPS) \cite{Toda2014} and Raman \cite {benhabib15}, $T_{\rm CDW}$ from X-ray \cite{SilvaNeto2014}.}. Remarkably, we find universal trends in the doping dependencies of the above energy scales. The PG, AN SC and the CDW energy scales decrease linearly as $p$ increases on a substantial doping range (panels (a)-(d)) in all these cuprate families. The $2\Delta_{\rm PG}$ scale, is about twice as large as that of the $2\Delta{\rm ^{AN}}_{\rm SC}$ and $2 \Delta_{\rm CDW}$ scales which are found to be very close to each other. On the contrary, the $2\Delta{\rm ^{N}}_{\rm SC}$ scale is non monotonic, it increases with doping up to the optimal doping level ($p=0.16$)(see panels (e)-(h)) and then it decreases in the over-doped regime ($p\geq0.16$). As a result, the $2\Delta{\rm ^{N}}_{\rm SC}(p)$ has a dome like shape fully observed in panel (g) and (h) for Hg-1201 and Bi-2212. If we now venture into a comparison between the doping dependence of these energy scales and the $T-p$ cuprate phase diagram (see panels (i)-(l)), the salient experimental facts are that $2\Delta_{\rm PG} (p)$ and the $2\Delta{\rm ^{N}}_{\rm SC}(p)$ follow the same behavior as $\Ts(p)$ and \Tc (p) respectively,  while $2 \Delta_{\rm CDW}(p)$ and $2\Delta{\rm ^{AN}}_{\rm SC}(p)$ do not follow  $T_{\rm CDW} (p)$ and $\Tc (p)$ respectively. At this stage, we are not in a position to propose a theory that would allow us to fully understand the doping dependencies of the energy scales and their correspondences with the characteristic temperatures of the cuprate phase diagram. However, if we focus on the doping dependency of the energy scales, we can draw some key observations from it. We can first hypothesize  that the three energy scales (PG, AN-SC and CDW) have probably a common microscopic origin since they have the same doping dependence. They decrease monotonically with doping as expected e.g. for the singlet formation energy in the resonant valence bound (RVB) model \cite{Anderson87}. Their microscopic origin could be e.g. short range anti-ferromagnetic fluctuations \cite{Scalapino1995} which decrease as one moves away by doping from Mott insulating anti-ferromagnetic phase \cite{Kotliar88,Bulut1994,Kyung04,Kyung2006,Kyung2009,Gull2013,Wu2017,Sordi2017,Wu2018}. We can also reasonably say that since $2\Delta_{\rm PG} (p)$ is at least twice as large as that $2\Delta{\rm ^{AN}}_{\rm SC}(p)$, it cannot be assigned to superconducting fluctuations as proposed by a preformed pair scenario \cite{Emery1995}. Another point worth mentioning is the same doping dependence of the N-SC and \Tc as opposed to the AN-SC gap that does not follow \Tc. This suggests that nodal quasi-particles are likely not subject to the same electronic interactions governing quasi-particles at the anti-nodes. On the other hand, the close values of the AN-SC and CDW energy scales, which we report here in several cuprates (panels (a)-(d)), is a surprising fact that deserves to be explored in the light of recent theoretical models of intertwined or composite orders\cite{Efetov2013,Sachdev13,Fradkin2014,Wang2015,Caprara2017,Chakraborty2019}. 
\section{Conclusion}
In conclusion, we have determined the universal energy scales behaviour associated with the  the $T-P$ cuprate phase diagram by extracting from electronic Raman scattering measurements the energy scales of the PG phase, of the anti-nodal and nodal superconducting state and of the charge density wave order for several cuprates families (Hg-1223, Hg-1201, Y-123 and Bi-2212). In all these cuprates, we find that $\Delta_{\rm PG}(p)$, $2\Delta^{AN}_{\rm SC}(p)$ and $2\Delta_{\rm CDW}(p)$ have the same doping dependence as $\Ts (p)$ : they decrease monotonically with doping. This suggests that they are all driven by the same microscopic interactions which could come for instance, from short range anti-ferromagnetic fluctuations. The closeness of the AN-SC and CDW energy scales suggests that these orders are intimately connected and that the pseudogap phase could be considered as intertwined or composite order of particle-particle and particle-hole pairs \cite{Efetov2013,Sachdev13,Fradkin2014,Wang2015,Caprara2017,Chakraborty2019}. On the contrary, the nodal component of the SC gap $2\Delta^N_{\rm SC}(p)$, which follows the same doping dependence as $\Tc (p)$, does not appear to be affected by any of the above interactions. Our experimental results will motivate future theoretical advancements that could account for these universal energy scales of the cuprate phase diagram.\\

\textbf{Acknowledgments} 
We thank the University of Paris, the Coll\`ege de France and the Canadian Institute for Advanced Research (CIFAR) for their support. 
B.L. was supported by the DIM OxyMORE, Ile de France. Work at Brookhaven is supported by the Office of Basic Energy Sciences, Division of Materials Sciences and Engineering, U.S. Department of Energy under Contract No. DE-SC0012704. Correspondence and request for materials should be addressed to A.S. (alain.sacuto@univ-paris-diderot.fr).

\appendix
\section{Details of the electronic Raman experiments}

Raman experiments have been carried out using a JY-T64000 spectrometer in single grating configuration using a 600 grooves/mm grating and a Thorlabs NF533-17 notch filter to block the stray light. The spectrometer is equipped with a nitrogen cooled back illuminated 2048x512 CCD detector. We use the 532 nm excitation line from a diode pump solid state with laser power maintained at 4 mW. Measurements between 10 and 290 K have been performed using an ARS closed-cycle He cryostat. This configuration allows us to cover a wide spectral range ($90~cm^{-1}$ to $2500~cm^{-1}$) with a resolution sets at $5~cm^{-1}$.  Spectra have been obtained from a single frame. Each frame is repeated twice to eliminate cosmic spikes and acquisition time is about 20 mn. All the spectra have been corrected for the Bose factor and the instrumental spectral response. They are thus proportional to the imaginary part of the Raman response function $\chi^{\prime \prime}(\omega,T)$. The B$_{1g}$ symmetry is obtained from crossed polarizations along the Cu-O bond directions. Then, the crystal is rotated by 45$\arcdeg$ using a Attocube piezo-rotator ANR 101 to obtain the B$_{2g}$ symmetry always using crossed polarizations. The B$_{1g}$ symmetry probes mostly the  principal axes of the BZ, (anti-nodal region) and it corresponds to the maximum amplitude of the SC gap  while the B$_{2g}$ symmetry probes mainly the diagonal of the BZ, (nodal region) and it corresponds to the region where the amplitude of the d-wave SC gap vanishes.
\section{Details on the crystal growth, doping and critical temperature of Hg-1223 and Bi-2212 single crystals}

\subsection{Hg-1223}

The Hg-$1223$ single crystals were grown by a single step synthesis \cite{Loret2017}. The as-grown single crystal has a critical temperature, \Tc $\approx$ 110 K. \Tc has been changed by annealing the single crystal under vacuum or oxygen. A thorough X-ray diffraction analysis reveals that oxygen atoms are removed (for under-doping) or added (for over-doping) inside the Hg layer \cite{Bertinotti1997}. The doping levels were estimated from the empirical Presland-Tallon's law \cite{Presland91}. 
\begin{figure}[ht!]
\begin{center}
\includegraphics[scale=0.45]{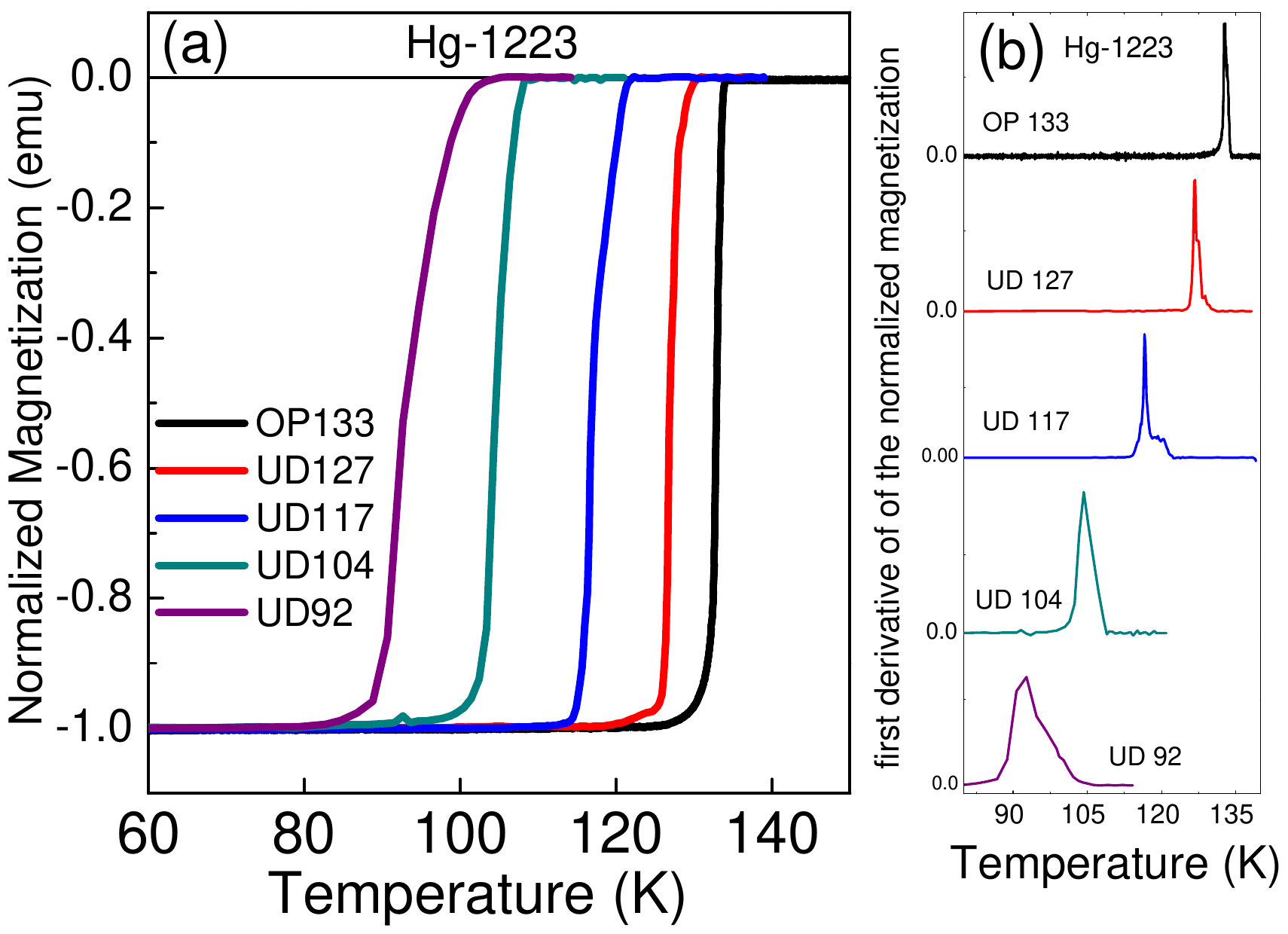}
\caption{(a) Zero field cooling magnetization curves of Hg-1223 single crystals for several doping levels. The applied magnetic field is perpendicular to the (ab) plane and its magnitude is of $\approx 10~Oe$.(b) First derivative of the magnetization curves displayed in (a). The location of the peak maximum indicates the value of \Tc and its full width at half maximum the transition width.}
\label{fig4}
\end{center}\vspace{-5mm}
\end{figure}
The single crystals are parallelepiped with a typical cross section of 0.7 $\times$ 0.7 mm$^2$ and a thickness of 0.2 mm. The c-axis is normal to the surface with the a-b plane directions 45$\arcdeg$ from the edges. In order to have high optical quality surface, the crystals have been polished using diamond paste at $1/10$ \micro m. Dc magnetization measurements under zero field cooling (ZFC) have been performed after polishing and displayed in Fig.~\ref{fig4} (a). The transition temperature \Tc and its width, $\Delta T_c$, was estimated by taking the maximum and the full width at half maximum of the peak of the first derivative of each Dc magnetization curve shown in Fig.~\ref{fig4} (b). The \Tc and $\Delta T_c$ values for each doping are the following: p=0.16 (\Tc= 133 K, $\Delta T_c$ = 1 K), p=0.14 (\Tc= 127 K,$\Delta T_c$ = 1.5 K), p=0.12 (\Tc= 117 K ,$\Delta T_c$ = 5 K), p=0.11 (\Tc=105 K,$\Delta T_c$ = 4 K), p=0.94 (\Tc= 92 K, $\Delta T_c$ = 7 K).  $\Delta T_c$ broadens when we move away from the optimal doping level. This reflects slightly doping inhomogeneity in the single crystal with under-doping. \\ 

\subsection{Bi-2212}

The Bi-2212 single crystals were grown by using a floating zone method. The optimal doped sample with $T_{c} = 90~K$ was grown at a velocity of 0.2 mm per hour in air ~\cite{Wen2008}. In order to get over-doped samples down to $T_{c}=65~K$, the as-grown single crystal was put into a high oxygen pressured cell between $1000$ and $2000$ bars and then was annealed from $350^{o}C$ to $500^{o}C$ during 3 days ~\cite{Mihaly1993}. The over-doped samples below $T_{c}=60~K$ was obtained from as-grown Bi-2212 single crystals put into a pressure cell (Autoclave France) with $100$ bars oxygen pressure and annealed from $9$ to $12$ days at $350~^{o}C$. Then the samples were rapidly cooled down to room temperature by maintaining a pressure of $100$ bars. The critical temperature $T_{c}$ for each crystal has been
determined from magnetization susceptibility measurements at a $10$ Gauss field parallel to the c-axis of the crystal. In the over-doped regime, $T_{c}$ increases linearly with $2\Delta^{AN}_{SC}$. From a linear fit of the $T_c$ values between $T_c=50\,K$ and $T_c=90\,K$, we find the reliable relationship:  $T_{c}=(2\Delta^{AN}_{SC})/8.2+28.6$ \cite{benhabib15}. In the under-doped regime $T_{c}$ falls down abruptly as a function of $2\Delta^{AN}_{SC}$ (see Fig.~\ref{fig5}). The level of doping $p$ was defined from $T_c$ using Presland and Tallon's equation \cite{Presland91}: $1-T_{c}/T_{c}^{max} = 82.6 (p-0.16)^{2}$. In the over-doped regime, estimate of $p$ can be determined from $2\Delta^{AN}_{SC}$ using the above two equations.
\section{Connection between the dip structure in the superconducting state and the normal state pseudogap}

\begin{figure}[ht!]
\begin{center}
\includegraphics[scale=0.6]{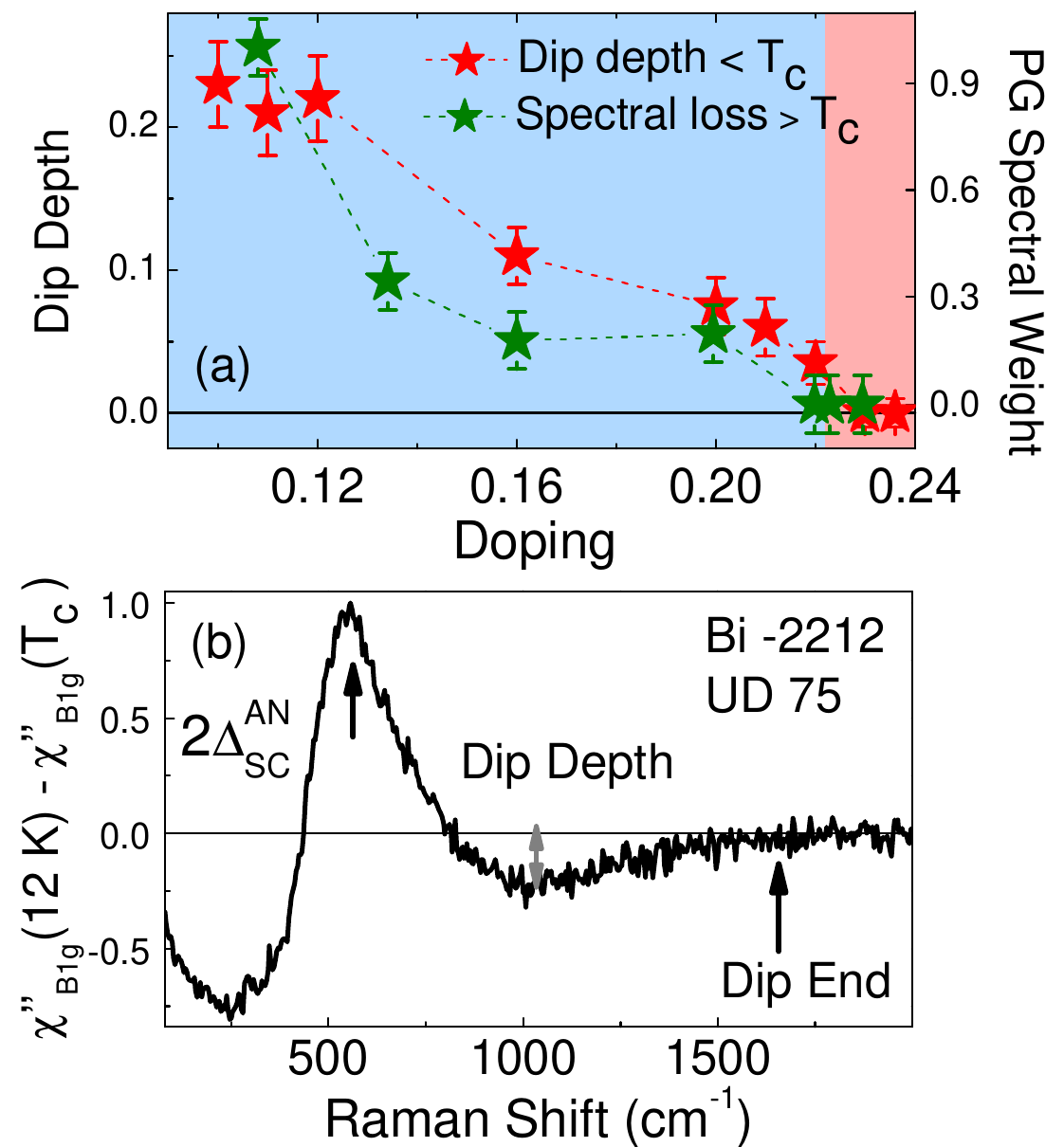}
\caption{(Color online). (a) Characteristic peak-dip structure extracted from the Subtracted Raman response of Bi-2212 (UD 75) single crystal between the SC and the normal state just above \Tc; (b) Doping evolution of the dip depth and the loss of spectral weight generated by the pseudogap phase.}  
\label{fig5}
\end{center}\vspace{-7mm}
\end{figure}

In order to show that there is a direct link between the dip-structure detected in the SC state Raman response and the spectral weight loss detected in the normal state Raman response when the pseudogap phase settles down, we have simultaneously plotted (see Fig.~\ref{fig5} (a)) the doping evolution of the dip and the loss of spectral weight in the Raman spectra of Bi-2212. The characteristic elements of the peak-dip structure measured on Bi-2212 (UD 75) are defined in Fig.~\ref{fig5} (b). 
We quantified the dip depth from the subtracted Raman response measured at low temperature ($\approx 12 $ K) in the SC state and just above \Tc. The loss of spectral weight is defined in ref. \cite{benhabib15}. From  Fig.~\ref{fig5} (a), it clearly appears that the dip depth and the loss of spectral weight are associated to the pseudogap phase. 
These results are supported by cellular dynamical mean field theory calculations (see appendix in ref.\cite{Loret2017a}).

\section{Extraction of the energy scales from the Raman response of Hg-1223 systems} 

\subsection{Superconducting anti-nodal and pseudogap energy scales versus doping}

\begin{figure} [ht!]
\begin{center}
\includegraphics[scale=0.27]{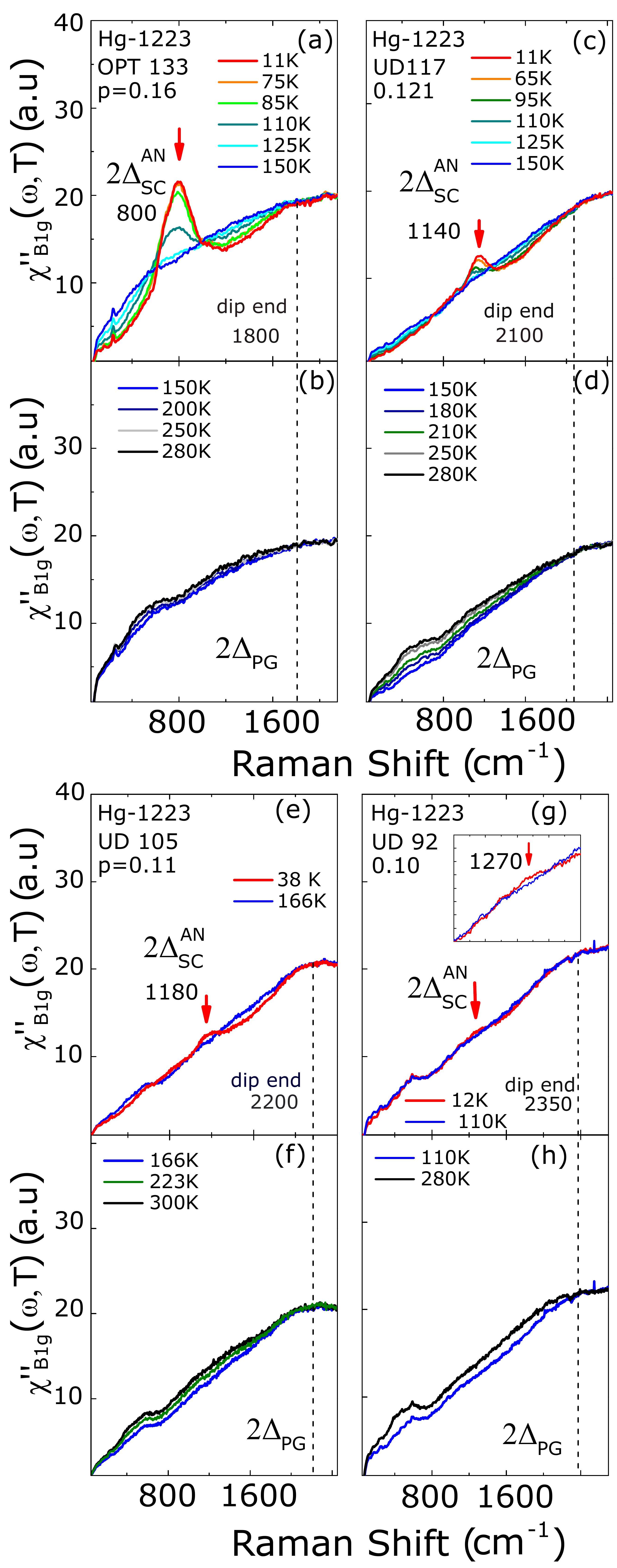}
\caption{(a),(c) and (e),(g): Temperature dependence of the \BAN Raman response function of Hg-1223 (with distinct doping levels) up to \Tc. (b),(d) and (f),(h): Temperature dependence of the \BAN Raman response function of Hg-1223 (with distinct doping levels) above \Tc. The pair breaking peak indicated by a red arrow determine the $2\Delta^{AN}_{\rm SC}$ scale while the pseudogap $2\Delta_{\rm PG}$ scale is defined from the energy for which the dip in the continuum ends. The inset in panel (d) corresponds to a zoom  of the \BAN Raman response in order to point out the $2\Delta^{AN} _{\rm SC}$. }
\label{fig6}
\end{center}\vspace{-5mm}
\end{figure}

The $\Delta^{AN}_{\rm SC}$ scale is the maximum energy of the $d-$wave SC gap which takes place in the anti-nodal region of the BZ. It is therefore experimentally observed in the \BAN Raman response. In fact, we are detecting twice the superconducting gap energy $2\Delta^{AN}_{\rm SC}$ which corresponds to the frequency of the pair breaking peak indicated by a red arrow in the top panels (a),(c) and (e),(g) of Fig.~\ref{fig6}. 
We see that $2\Delta_{\rm SC}$ decreases in intensity and increases in frequency as $p$ is lowering from 0.16 to 0.10. This doping dependence is a common feature to all the cuprates studied (cf. Fig. 3). Its rapid intensity decrease is likely due the loss of the spectral weight in the anti-nodal region generated by the PG phase. On the other hand, the PG energy scale, $2\Delta_{\rm PG}$ is defined as the frequency for which the dip just on the right side of the pair breaking peak ends. Remarkably, it is approximately at the same frequency than the one for which the PG depletion ends in the normal state (see panels (b), (d) and (f), (h)). This is pointed out by the dashed line in the pairs of panels (a,b), (c,d) and (e,f) (g,h) of Fig.~\ref{fig6}. Note that this is not always the case. 

\subsection {Nodal superconducting and charge density wave energy scales} 
\begin{figure}[hb!]
\begin{center}
\includegraphics[scale=0.27]{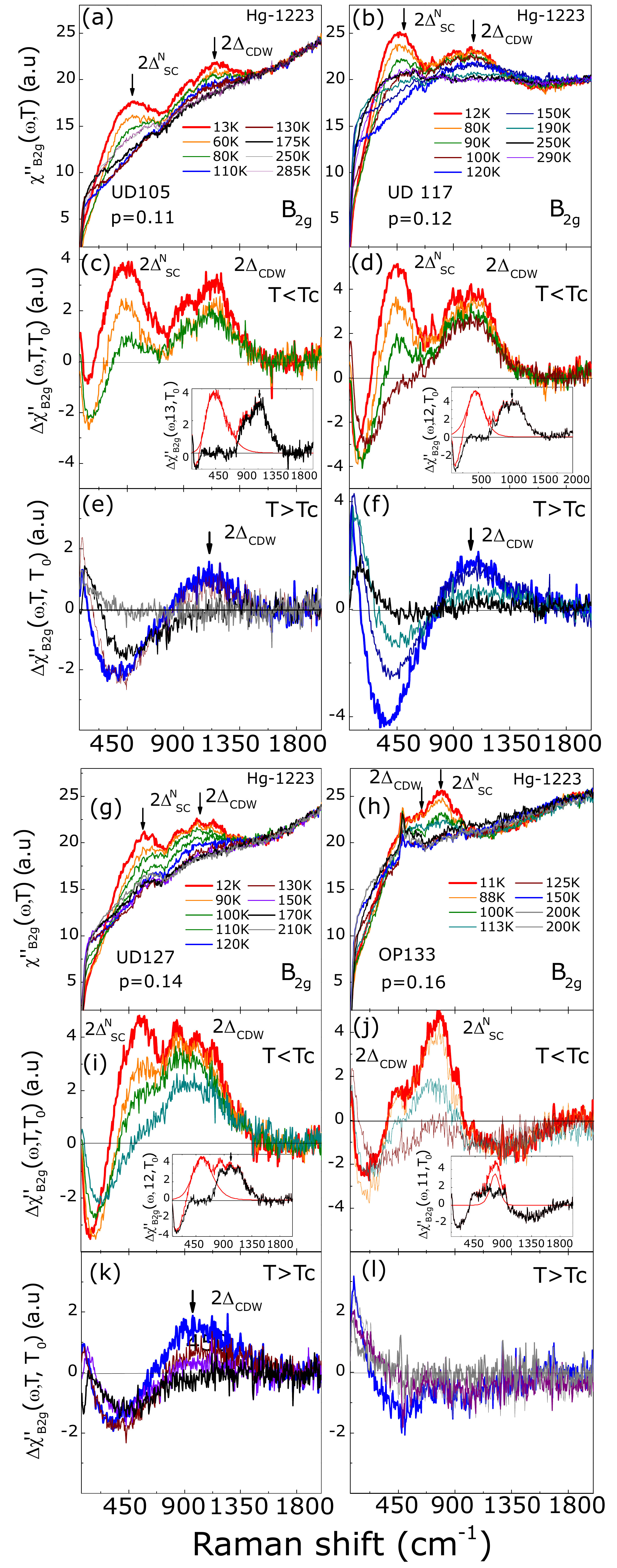}
\caption{(Color online). (a), (b) and (g), (h): Temperature dependence of the nodal Raman responses (\BN) of HgBa$_2$Ca$_2$Cu$_3$O$_{8+\delta}$ (Hg-1223) for several doping levels. The features related to the CDW and the nodal SC gap are indicated by black arrows. (c),(d) and (h), (j)  : Nodal Raman responses below \Tc, after subtracting the one at $T_{0}$. The $T_{0}$ values for each doping are listed in the text. In the insets, the black curve corresponds to the CDW hump rid of the nodal SC component, the full red curve is a ASG fit of the SC nodal gap subtracted (see text for more details). (e),(f) and (k),(l): Nodal Raman responses above \Tc, after subtracting the one at $T_{0}$ to highlight the CDW structure (dip and hump).}  
\label{fig7}
\end{center}\vspace{-7mm}
\end{figure}

The temperature dependence of the \BN Raman responses of the Hg-1223 compound for several doping levels are shown in Fig.~\ref{fig7}. In the panels (a), (b) and (g), (h) of Fig.~\ref{fig7}, we detect  both  $2 \Delta_{\rm CDW}$ and the nodal SC gap $2\Delta{\rm ^N}_{\rm SC}$. 

\begin{figure}[ht!]
\begin{center}
\includegraphics[scale=0.55]{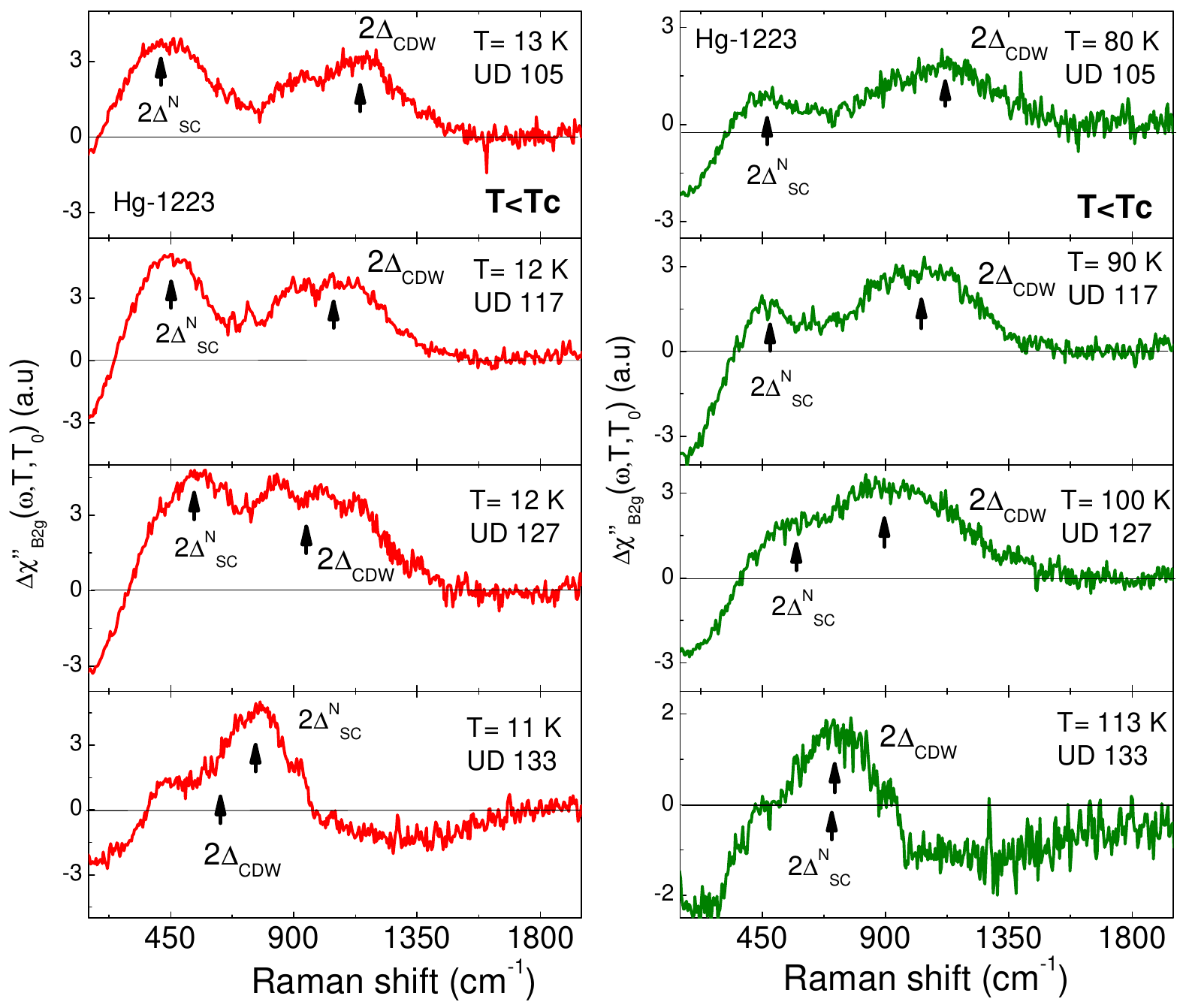}
\caption{(Color online). Left and right panels are respectively the subtracted Raman responses of Hg-1223, $\Delta\chi^{\prime \prime}_{\BN} (T \approx12\,K,T_{0}$) and $\Delta\chi^{\prime \prime}_{\BN} (\Tc\approx25\, K,T_{0}$) for various doping levels. The $T_{0}$ values for each doping are listed in the text. The black arrows indicate the location of $\Delta_{\rm CDW} (p)$ and $\Delta^N_{\rm SC}(p)$. The red dotted line is just a guide for the eyes.}
\label{fig8}
\end{center}\vspace{-7mm}
\end{figure}

For $p=0.11$, $2 \Delta_{\rm CDW}$ and $2\Delta{\rm ^N}_{\rm SC}$ are well separated in frequency. However, as $p$ increases, they are getting closer in frequency and for $p=0.16$, they are almost superimposed. In order to stress these two gaps, we plotted $\Delta\chi^{\prime \prime}_{\BN} (\omega,T,T_0) =\chi^{\prime \prime}_{\BN} (\omega, T)- \chi^{\prime \prime}_{\BN} (\omega, T_0$) where $T_0$ takes the values: 285 K, 290 K , 210 K 280 K for respectively UD 105, UD 117, UD 127 and OP 133 (see panels (c),(d)) and (h),(j)).  As $T$ increases up to \Tc , the intensity of the nodal component of the SC gap is strongly reduced while the intensity of the CDW hump remains almost constant (see black arrows). We can bring out the CDW signal below \Tc by taking off the SC nodal gap contribution after fitting it by an asymmetric Gaussian (AsG) function (see insets in Fig.~\ref{fig7}). The set of the fitting parameters used for the doping levels $p$= 0.11, 0.12,0.14 and 0.16 (at $T\approx 12$ K) are respectively (A=5, $\omega_c$=441~\cm, $\omega_1$=292~\cm, $\omega_2$=47~\cm, $\omega_3$=116~\cm), (A=13, $\omega_c$=456~\cm, $\omega_1$=80~\cm, $\omega_2$=90~\cm, $\omega_3$=85~\cm),(A=8, $\omega_c$=540~\cm, $\omega_1$=200~\cm, $\omega_2$=80~\cm, $\omega_3$=110~\cm) and (A=7, $\omega_c$=770~\cm, $\omega_1$=80~\cm, $\omega_2$=50~\cm, $\omega_3$=40~\cm). Above \Tc, the nodal SC gap, $2\Delta{\rm ^N}_{\rm SC}$, is gone and only remains the CDW gap: a dip-hump structure (see $\Delta\chi^{\prime \prime}_{\BN} (\omega, T,T_0)$ in panels ((e),(f) and (k), (l)). Note that the CDW dip-hump structure is observable in the Raman spectra for $p$=0.11, 0.12 and 0.14 while for $p$=0.16 is hardly detectable, likely because the CDW signal collapses below or close to \Tc = 133 K. We can improve the determination of the $2\Delta_{\rm CDW}(p)$ value by analyzing the nodal Raman responses at low temperature. However, the extraction of $2\Delta_{\rm CDW}(p)$ is complicated by the existence of SC signal. See the subtracted Raman response $\Delta\chi^{\prime \prime}_{\BN} (\omega, T\approx 12 K, T_0)$ of Hg-1223 (left panel of Fig.~\ref{fig8}). The situation is even more complex for $p$=0.16 where the Raman CDW signal coincides with the nodal SC one. Yet, if we increase $T$ but stay below \Tc, this allows us to weaken the SC signal and bring out the CDW signal and obtain reliable $2\Delta_{\rm CDW} (p)$ values. This is achieved by measuring  $\Delta\chi^{\prime \prime}_{\BN} (\omega,T,T_0)$  with $T \approx$ 25 K  below \Tc for each doping level (see right panel of Fig.~\ref{fig8}). We then find that $2\Delta_{\rm CDW}$  = 1150 \cm, 1030 \cm, 930 \cm and 730 \cm for respectively $p$=0.11, 0.12; 0.14 and $p$=0.16. Note that this procedure is applicable because as it can be seen in Fig.~\ref{fig7}, the location of CDW hump is almost temperature independent.

\clearpage

\bibliographystyle{apsrev4-1} 
\bibliography{cuprates}

\end{document}